\documentstyle{l-aa}

\begin{document}

\thesaurus { 11.07.1, 11.09.2, 11.16.1, 11.16.2, 11.19.2 }

\title{ Global structure and formation of polar-ring galaxies}

\author{V.~Reshetnikov\inst{1,2}, N.~Sotnikova\inst{1}}

\offprints{N.~Sotnikova}

\institute{
           Astronomical Institute of St.Petersburg State University,
           198904 St.Petersburg, Russia
\and
            DEMIRM, Observatoire de Paris, 61 Av. de l'Observatoire,
 F-75014 Paris, France}

\date{Received 1996; accepted}

\maketitle

\begin{abstract}
We present an analysis of structural features of all known galaxies 
with optical polar rings. 
We find a clear dichotomy for objects of this peculiar 
class. Bulge-dominated S0 galaxies possess only short 
narrow rings, while disk-dominated objects always have wide 
extended polar rings. We try by gas dynamical simulations 
to explain such a segregation by dependence of
the ring-forming 
process on different galaxy potentials. It is found that the 
total mass captured into the ring during an encounter of a 
host-ring system with a gas-rich spiral galaxy of comparable 
mass exceeds $10^9 M_{\odot}$ (or about 10\% of all gas in the
donor galaxy), which is of the order of 
that found by observation. The process of gas to gather 
into a steady-state ring takes approximately $(7 - 9)\times10^8$ 
years. This time is somewhat shorter for rings forming around 
bulge-dominated galaxies. We also present observational arguments 
for S0 galaxies with extended rings to be similar to late-type 
spirals by their photometric properties, while numerical modelling 
of the extended ring formation suggests that these 
galaxies must possess massive dark halos as well. In this case, 
the sizes of the modelled rings turn out large enough (up to 30~ kpc 
in diameter), and the time scale for ring formation is prolonged 
up to several Gyrs.


\end{abstract}

\section{Introduction}

Recent observations (and especially with HST) demonstrate
that mutual interactions and mergers between galaxies at early
stages of evolution of the Universe were probably among the main
processes leading to the observed properties of galaxies
(e.g., Keel 1996). Even at the present epoch, at least 5-10\% of
galaxies are members of interacting systems. Many other
galaxies keep signs in their structure of past interactions
and mergings (for example, elliptical and S0 galaxies with inclined
gaseous disks, galaxies with faint shells and ripples, galaxies with
kinematically decoupled nuclei, etc.). 

Polar-ring galaxies (PRGs), 
consisting of large-scale rings of stars, gas and dust orbiting around
major axes of early-type galaxies, may be considered as extreme samples
of possible interaction relics. Indeed, in the case of PRGs, 
the remnants of merged galaxies are not mixed in one smooth object 
but stay separately in a quasi-steady state for a long time.
PRGs are very rare objects. For example, the Polar Ring Catalogue by
Whitmore et al. (1990) (PRC) lists only 6 classic kinematically-confirmed
polar-ring galaxies. 

The unique geometry of PRGs attracts the attention of astronomers trying
to test the 3D shape of galactic potentials and to study the stability of 
inclined rings and disks (see recent review articles by Tohline 1990, 
Sackett 1991,
Sparke 1991, Combes 1994, Cox \& Sparke 1996). Such an important 
question
as the origin of these peculiar objects it is still not 
adequately investigated. 
It is usually suggested that collapse of a single protogalactic cloud
cannot create an object with two nearly-orthogonal large-scale systems 
(but see Curir \& Diaferio 1994), 
and so some secondary event must occur in the history of PRGs. 

Summarizing possible scenarios of polar-ring formation, one can 
enumerate the following:
the accretion of matter from a nearby system or 
the capture and merging of a gas-rich companion;
the delayed inflow of a primordial intergalactic cloud;
the accretion of matter from the outskirts of the host galaxy itself; 
the polar-ring formation from the return of the tidal material during 
the merging of two gas-rich spirals 
(Toomre 1977, Shane 1980, Schweizer et al. 1983,
Sackett 1991, Sparke 1991, Hibbard \& Mihos 1995).

Probably, all the above mechanisms can create ring-like 
structures around early-type galaxies. To our mind, 
the most straightforward scenario is the first one. Recent
observations of several binary interacting systems clearly
demonstrate such rings in the making (for instance, NGC 7464/65 - 
Li \& Seaquist 1994, 
NGC 3808A,B and NGC 6285/86 - Reshetnikov et al. 1996).
 
Another unclear question is the nature of central objects in PRGs
and a possible correlation of host galaxy properties with
characteristics of a surrounding polar ring. 

In this paper, we present SPH simulations of polar ring formation 
around target galaxies of different structures due to gas 
accretion during the encounter
with a comparable-mass spiral galaxy. In our simulations, we
try to answer the following main questions: 
Does this mechanism work? 
What determines the size of the resulting ring, and what is its 
spatial structure? 
On what timescale does the ring form?
What is the mass fraction of the gas captured into a ring? 
      
The paper is organized as follows: in Section 2, we examine
observational properties of all known kinematically confirmed
PRGs and formulate some observational constraints on numerical 
simulations; in Section 3, we discuss previous attempts
to model the PRGs formation, describe our modelling technique and
results of simulations; and finally we give our conclusionsin Section 4.

Throughout the paper, all distance-dependent quantities are
calculated using $H_{0}~=~$75 km/s/Mpc.

\section{General characteristics of polar-ring galaxies}

As a definition of a polar-ring galaxy, we will use the definition
of Category A objects in the PRC: spectroscopic evidence must
exist for two nearly-perpendicular kinematical subsystems;
centers of the two components must be aligned, and 
both subsystems must have similar systemic velocities;
the ring must be comparable in size to the host galaxy, must
be luminous and nearly planar. This definition allows to separate
dust-lane ellipticals, galaxies with inclined HI rings etc.
from PRGs.
Using this rigorous definition, one can now consider only three 
additional galaxies to 6 the classic PRGs listed in the PRC:
AM 2020-504 (Whitmore \& Schweizer 1987, Arnaboldi et al. 1993),
IC 1689 (Reshetnikov et al. 1995, Hagen-Thorn \& Reshetnikov 1997)
and NGC 5122 (Cox et al. 1995). (We do not consider ESO 603-G21 
here due to the puzzling kinematics of the central galaxy (Arnaboldi
et al. 1995).)

An examination of the optical images of PRGs (e.g. in the PRC)
allows one to divide them into two groups (Whitmore 1991):
galaxies with extended disk-like rings with the central
region cut out and galaxies with relatively narrow rings, not 
extended in radius. This division is quite distinct
since the first group of galaxies - A0136-0801 (A-1), UGC 7576 (A-4),
NGC 4650A (A-5), UGC 9796 (A-6), and NGC 5122 (B-16) - possess
optical rings extended out to 2-3 diameters of the central
galaxies, while the second group - ESO 415-G26 (A-2),
NGC 2685 (A-3), IC 1689 (B-3), and AM 2020-504 (B-19) -
demonstrate optical rings with size not exceeding 
the diameter of the host galaxy. 

In Table~ 1, we generalize the main observational characteristics
of the two groups of PRGs. (Note that, due to the absence of 
optical data about NGC 5122, we did not consider this galaxy
in the table.) 
In the case of incomplete data or large scatter of
characteristics, we give in the table only limits or
indicate the range of parameter changes. 
Absolute luminosities and
colors in the table are corrected for Galactic absorption.  

We discuss Table~ 1 in detail.

\begin{table*}
\caption[1]{General characteristics of PRGs}
\begin{tabular}{lll}
\\
\hline
Parameter & Extended rings & Short rings \\
\hline \\ 

{\it Central galaxy}: \\ 
Bulge-to-disk ratio & $\rm <<1$  &  $\approx$1 \\ 
Absolute luminosity in the $B$ band & $\rm -18.6~\pm~0.5(\sigma)$ & $\rm
-19.2~\pm~0.3(\sigma)$ \\ 
$B-V$    & $\rm +0.84~\pm~0.05$ & $\rm +0.91~\pm~0.03$ \\ 
Colour gradient & Yes & Yes \\ 
$\sigma$ (km/s) & $\rm 80~\pm~20$ & $100 \div 270$ \\ 
$\rm V_{max}$/$\sigma$ & $\rm 1.7~\pm~0.3$ & $0.45 \div 1.4$ \\ 
Decoupled nuclei &  ? & Yes \\ \\

{\it Ring}:\\
Diameter        & $\rm \approx~(2 - 3)D_{25}$ & $\rm \leq~D_{25}$ \\
                & $\rm 26~\pm~6$ kpc &  $\rm 9~\pm~4$ kpc         \\ 
Absolute luminosity in the $B$ band   & $\rm -17.5~\pm~0.6$ & $\rm \geq~-15.4$ \\
$B-V$  &  $\rm -0.1~to~+0.7$ &  $\rm +0.2~to~+0.64$ \\
Colour gradient & Yes & ?  \\
HI mass ($10^{9}\,M_{\odot}$) & $\rm 6~\pm~2$  & $\rm \leq~9$ \\ 
Ring inclination & $\rm 7^{o} \div 26^{o}$ & $\rm 1^{o} \div 16^{o}$ \\ \\
\hline
\end{tabular}
\end{table*}

\subsection{Central galaxies}

The problem of the bulge-to-disk ratio determination for the PRGs is not
quite simple. Reshetnikov et al. (1994) have noticed that bulge effective
parameters ($\mu_{e}$ and $\rm R_{e}$) of PRGs with 
wide extended rings (our first group) lie in the plane of
effective parameters below the mean relationship for normal
galaxies (this means that at the same value of $\mu_{e}$, bulges
of PRGs have smaller radii in comparison with bulges of normal galaxies) -
see Figure 7 in the cited paper. 
But bulge characteristics of the second group of galaxies are
located exactly along the standard relationship. Such a dichotomy
suggests two explanations: the unusual compactness of the first
group of PRG bulges or the underestimation of their sizes and
luminosities due to
the projection of gas and dust rich extended rings on central
regions of the galaxies (Reshetnikov et al. 1994).

A comparison of photometric cuts of the galaxies in different
color bands allows one to solve this problem. In Figure~\ref{Fcuts}, 
we present
surface brightness distributions for four PRGs with extended
rings along the major axes of their central galaxies. We compare
two profiles for each galaxy: one in the $B$
passband and other in the red filter ($R$, $i$, or $K$).
As one can see from figure, absorption in the rings barely
changes the shape of the profiles. Therefore, PRGs with wide
extended rings in fact possess unusually compact and faint bulges
in comparison with normal early-type galaxies. As is evident 
from Figure~\ref{Fcuts}, 
the exponential disk dominates in the photometric structure of all these 
galaxies. The ratio of the total bulge luminosity to the luminosity
of the exponential disk in the first group of PRGs is about 0.1 (we
denoted this as $\rm <<1$ in Table~ 1).

\begin{figure*}
\picplace{0.8cm}
\caption[1]{Photometric profiles of PRGs with extended rings
along major axes of the central galaxies.
The data for A0136-0801 are from Schweizer et al. (1983) and PRC;
NGC 4650A $-$ Whitmore et al. (1987) and Combes \& Arnaboldi (1996);
UGC 7576 and UGC 9796 $-$ Reshetnikov et al. (1994). The dashed
lines represent exponential fits of the observed distributions.} 
\label{Fcuts}
\end{figure*}  

The photometric structure of PRGs with short rings (the second
group of galaxies) looks usual for early-type galaxies. As
was mentioned earlier, their characteristics in the
plane of effective parameters follow the mean relation for
normal galaxies. According to original papers, 
bulge-to-disk ratios in the $B$ band for these galaxies are:
ESO 415-G26 $-$ 0.52 (Whitmore et al. 1987), NGC 2685 $-$ 0.9
(Makarov et al. 1989), IC 1689 $-$ 1.9 (Reshetnikov et al. 1995).
AM 2020-504 is an elliptical galaxy (Arnaboldi et al. 1993).
Therefore, the second group PRGs are normal bulge-dominated galaxies
by their photometric structure (conditionally, $\rm \approx 1$ in
the table).

\subsection{Polar rings}

Rings sizes in the two groups of galaxies are different,  
as well as their absolute luminosities ($\rm D_{25}$ in Table~ 1 
denotes a diameter of the central galaxy measured at the surface brightness 
level $\mu_{B}\,=\,$25). 
Both groups of rings show a large scatter 
of optical colors although, as it was shown by 
Reshetnikov et al. (1995), there is a general trend with a blue 
surface brightness; bluer rings have, on average, 
higher surface brightnesses. Extended rings also show large-scale 
color gradients, that is, they become bluer at larger radii 
(see, for instance,  Figures 4,6 in Reshetnikov et al. 1994 and 
Arnaboldi et al. 1995). Both group of rings contain a large amount 
of neutral hydrogen (we assume in Table~ 1 that all detected HI 
belongs to the rings), with larger scatter of HI mass in PRGs with 
narrow rings (Schechter et al. 1984, van Gorkom et al. 1987, 
Richter et al. 1994). The last line of Table~ 1 shows the range of ring 
inclinations (angular distance between the ring and the perpendicular
to the central galaxy plane) according to Whitmore 
(1991). Both groups of PRGs demonstrate rings to be very close to 
perpendicular, with a somewhat larger deviation for extended rings.
It should also be noted that global characteristics of extended
rings resemble the disks of spiral galaxies (this was
previously pointed out by Reshetnikov et al. 1994 and Reshetnikov \&
Combes 1994a,b). We will discuss this analogy in Section 4.  

Summarizing the above analysis, one can conclude that
{\it there is a correlation between general properties
of optical polar rings and characteristics of host galaxies.}
Extended, disk-like rings exist preferably around galaxies
which have global photometric structure (and probably mass
distribution) similar to late-type spiral galaxies. Host
galaxies of PRGs with extended rings demonstrate a remarkable
similarity $-$ their characteristics fall in a relatively
narrow range (see~ Table~ 1). From the other side, bulge-dominated
galaxies have tighter and narrow rings. Host galaxies and rings 
of PRGs with non-extended rings demonstrate significantly larger
scatter of intrinsic properties. 

Although the statistics of
PRGs are not great (8~-~9 objects only), the mentioned tendency
is quite distinct~ $-$~ there are no bulge-dominated galaxies
with extended luminous polar rings. It is significant since the
known PRGs were selected on the base of optical morphology only,
without any attention to the structure of the central galaxies.
One can note also that
the PRG dichotomy is quite analogous to a recently-found
difference of ionized gas distribution in elliptical and S0
galaxies. According to Macchetto et al. (1996), more than half
of elliptical galaxies with detected ionized gas have their gas concentrated
in small ($\leq$4 kpc) nuclear disks, while most S0 galaxies
demonstrate more extended (up to 18 kpc) distribution of the gas.   

A similar tendency was mentioned for the first time by Whitmore (1991),  
who found that only rapidly-rotating S0 disks have 
extended luminous polar rings, while dust-lane ellipticals rarely
show an extended luminous component. He suggested that a more
flattened potential of the S0 galaxy is able to stabilize
the ring at greater radii than in an elliptical galaxy. 
In the next Section, we will demonstrate that the observed dichotomy 
could be caused by another reason.
 
\section{Numerical simulations}

There are only a few  
investigations that illustrate several proposed scenarios of 
polar ring formation. 
This lack of theoretical study devoted 
to the question of ring origin is connected with the necessity 
of the usage of complicated 3D gas dynamics models, since polar
rings contain a high amount of neutral gas (up to a few 
$10^9 \, M_{\odot}$), as seen from Table~ 1. 
One of the 
most fit tools to construct such models is a smoothed particle 
hydrodynamics algorithm -- SPH, -- originally proposed by Lucy (1977) 
and Gingold \& Monaghan (1977) and considerably developed in the 
late eighties (e.g., Hernquist \& Katz 1989). It is fully 
three-dimensional, flexible regarding the geometry of the gas 
distribution and consequently well-suited to modelling both the 
donor and recipient galaxies and the empty extended space between them. 
Applied to a range of polar-ring investigation problems, this 
method has already lead to many important results. 

The main result of the ring stability investigations is that the 
time for gas in an axisymmetric or triaxial potential to settle 
into a steady state is small compared to the age of the Universe 
(see Christodoulou et al. 1992 and references therein). So there is 
sufficient time for accreted gas to form polar rings. However, 
these studies did not take into account time-dependent effects 
and realistic initial conditions. Some efforts to include these 
effects have been made by Rix \& Katz (1991) and 
Weil \& Hernquist (1993). 

Rix \& Katz (1991) have treated the polar-ring formation process as 
a gradual smearing of a small gaseous blob captured by a galaxy of 
spherically-symmetric structure on a distant circular orbit. Naturally, 
the diameter of the forming ring (up to 80 kpc) was quite completly 
dictated by the initial position of a gaseous satellite (the orbit 
radius). The approach of Weil \& Hernquist (1993) was more 
self-consistent. They have considered a parabolic encounter of a 
low-mass gas-rich companion with a more massive galaxy and consequent 
merging of the former. The gaseous component of the satellite was 
settling into the ring after the total disruption of the satellite 
in the vicinity of the target galaxy. The forming ring 
turned out rather small (about 6-7 kpc in diameter) and its size was 
determined in the end by a hand-introduced description of details of 
the companion disruption. 

The ability of the accreting material to dissipate energy 
is the crucial factor for ring formation. Test (noninteracting) 
particles (which imitate molecular clouds from intergalactic space 
or clouds belonging to a companion galaxy) lose 
energy due to ram-pressure in an extended halo of a spiral  
galaxy and form a gaseous ring rotating around the disk of the spiral 
(Sofue \& Wakamatsu 1993, Sofue 1994).

\subsection{Modelling of polar-ring characteristics 
during the ring-forming process}

The main feature of the present investigation is a description of 
the full history of the gas stripping of the spiral galaxy outskirts 
and its consequent capture by a satellite during a parabolic encounter. 
Keeping in mind the recent observations of several 
interacting pairs of galaxies of comparable luminosities which 
demonstrate evidently ring-like structures in the making (see Introduction), 
we have considered a distant encounter of equal-mass systems. 
In the case of a distant encounter the effect of self-gravity may be 
ignored. The treatment 
of the gas hydrodynamics is described below. 

We did not explore in a comprehensive manner 
the orbital parameter space and choose only one set of impact 
parameters for which the modelling encounter of an S0 galaxy with a 
gas-rich system of comparable mass unambiguously results in the polar-ring 
formation around the former. Motivated by the above observational 
data analysis, we have undertaken a numerical investigation  
of the ring-forming process for galaxies with 
different structures. 
 
\subsubsection{Method}
Our simulations are based on a rather standard variant of the SPH 
code, which is the same as that in Sotnikova (1996). For simplicity,  
we adopt the smoothing length $h$ (analogue of the particle size) to 
be independent of the local gas density. The smoothed values of 
hydrodynamical parameters are estimated by using a grid with  
cell length equal to $2h$. We assume an isothermal equation 
of state, leaving thermal effects completely unexplored. The 
gas is always at a temperature $10^4 \, {\rm K}$,  and the corresponding 
equation of state is
$$
P = c^2\rho \, ,
$$ 
where $P$ and $\rho$ are the gas pressure and density, and $c~=~const$ 
is the isothermal speed of sound. For $T~=~10^4 \,{\rm K}$ \, ,
$c\approx9.1 \, {\rm km/s}$.

The motion of $N = 10\,000$ particles, which represent elements of the 
continous gaseous medium, is traced by means of equations of momentum 
conservation:
\begin{eqnarray}
\frac{d \vec{r}_i}{d t} &=& \vec{v}_i \, ,
\label{coord} \\
\frac{d \vec{v}_i}{d t} &=& - \frac{1}{\rho(\vec{r}_i)} \vec{\nabla} P(\vec{r}_i) + 
\vec{a}^{\rm \, visc}_i - \vec{\nabla} \Phi(\vec{r}_i) \, ,
\label{vel} 
\end{eqnarray}
where $\vec{r}_i$ and $\vec{v}_i$ are the spatial coordinate and velocity 
of the $i$-th particle, and $\Phi$ is the gravitational potential. The term 
$\vec{a}^{\rm \, visc}_i$ describes the acceleration due to viscosity 
$q_i$:
$$
\vec{a}^{\rm \, visc}_i = - \frac{1}{\rho_i} \vec{\nabla} q_i \, ,
$$
where $\rho_i = \rho(\vec{r}_i)$. We adopt one of the standard forms for the 
artificial viscosity $q_i$ which serves to represent the sudden 
deceleration of the gas motion when strong shocks arise in the gas 
(e.g., Hernquist \& Katz 1989):

\begin{equation}
q_i = \left\{ 
\begin{array}{ll}
\alpha c h \rho_i |\vec{\nabla} \cdot \vec{v}_i| + 
\beta h^2 \rho_i |\vec{\nabla} \cdot \vec{v}_i|^2  
, & \vec{\nabla} \cdot \vec{v}_i \leq 0 \\
0 ,                                & {\rm otherwise} \, . 
\end{array}
\right.
\label{visc}
\end{equation}
In the expression (\ref{visc}), the viscosity depends on the divergence of 
the velocity field. The first term is analogous to a bulk viscosity, the 
second, indroduced to prevent particle interpenetration at high 
Mach number, is of the Neumann-Richtmyer type. Parameters $\alpha$ and 
$\beta$ are free parameters. According to Hernquist \& Katz (1989), one 
can satisfactorily reproduce the change of the density and pressure across 
the shock front if the values of $\alpha$ and $\beta$ are equal to 0.5 
and 1.0, respectively. We used just these numbers in our simulations.

The procedure of a smoothed-value estimation of the gas density is 
quite usual, as well as the transition from hydrodynamical equations 
(\ref{coord}) and (\ref{vel}) to SPH-equations (for a detailed 
description of the adopted numerical scheme, see Sotnikova 1996).  

\subsubsection{Model} 
{\it Donor galaxy}

To minimize the number of free parameters, we chose a very simple 
model for the donor galaxy -- its potential was taken as that of 
a softened point mass
\begin{equation}
\Phi_0(r) = - \frac{G\, M_0}{(r^2+a_0^2)^{1/2}} \, ,\\
\label{spiral}
\end{equation} 
where $G$ is the gravitational constant, $M_0$ the mass, and 
$a_0$ the softening scale length of the potential. 

$N = 10\,000$ particles are used to represent the gaseous medium. 
They are initially gathered in a thin disk with a $1/r$ distribution 
from the center of the donor galaxy up to an outward radius $R_0$ and 
are placed in dynamically-cold circular orbits. The total gas mass 
is 0.2 $M_0$. The gaseous particle size $h$, which 
determines the numerical spatial resolution, is 300 pc.

Throughout, we employ the following system of units: the gravitational 
constant $G = 1$, the mass of the donor galaxy $M_0 = 1$, the outer 
radius of the initial gas distribution in the donor galaxy $R_0 = 1$. 
In terms of physical units, a unit mass can be defined as 
$10^{11} M_{\odot}$, and a unit distance corresponds to 15 kpc. 
Combined together, they give a unit time to be 86.6 Myrs, 
and a unit velocity to 169.3 km ${\rm s}^{-1}$. \\ \\
{\it Accreting galaxy}

We have considered the two-component model of a galaxy, consisting 
of a bulge and a disk. The light 
distribution of the bulge is well-provided by 
the model used, for example, in Weil \& Hernquist (1993). For 
this model, the potential is
\begin{equation}
\Phi_{\rm b}(r) = - \frac{G\, M_{\rm b}}{r+a_{\rm b}} \, ,\\
\label{bulge}
\end{equation} 
where $M_{\rm b}$ is the bulge mass, and $a_{\rm b}$ the 
scale length of the potential. 

The gravitational potential of the disk component of the galaxy 
is approximated by a Miyamoto \& Nagai (1975) potential
\begin{equation}
\Phi_{\rm d}(x,y,z) = - \frac{G\,M_{\rm d}}{\sqrt{y^2 + z^2 + 
\left(b_{\rm d}+\sqrt{x^2 + a_{\rm d}^2}\right)^2}} \, ,\\
\label{disk}
\end{equation}
where $M_{\rm d}$ is the disk mass, and 
$a_{\rm d}$ and $b_{\rm d}$ are scale radii. 
This model was favored 
by the simplicity of its practical usage. In our simulations,  
$a_{\rm d}/b_{\rm d} = 0.2$.

The ratio $M_{\rm b}/M_{\rm d}$ determines the range of models 
from bulge-dominated systems (large values of the ratio) to 
disk-dominated galaxies (small values of the ratio). Four models 
for an accreting galaxy have been considered, with values of 
$M_{\rm b}/M_{\rm d}$ equal to: 2.0, 1.0, 0.5, 0.2. 
The total mass of the galaxy is taken the same for 
all models and equal to that of the donor galaxy: 
$M_{\rm b} + M_{\rm d} = 1 = 10^{11} M_{\odot}$.

As shown in Section 2, main galaxies of PRGs with extended 
rings are similar to spiral galaxies in global photometric 
structure (and probably mass distribution). 
According to de Jong (1996), the bulge effective surface brightness 
shows the best correlation 
with morphological type of spiral galaxies. Therefore, 
the total bulge luminosity (mass) also correlates well. 
Scaling 
radii show a large scatter for all types of spiral galaxies. 
Thus, we fixed the scale lengths 
of galaxy components, leaving only the mass of the bulge to be 
variable. Scale lengths of the bulge $a_{\rm b}$ and disk $b_{\rm d}$ 
potentials are 1 kpc and 5 kpc, respectively. For these values, the 
half-mass radius of the bulge is $(1+\sqrt{2})a_{\rm b}\simeq 2.4$ kpc, 
and the maximum of the rotation curve of the disk component falls on 
the radius $\sqrt{2}(b_{\rm d}+a_{\rm d})\simeq8.4$ kpc. \\ \\
\noindent{\it Orbit}

Before modelling a ring-forming encounter, we first solved the 
two-body problem for the donor and ring-host galaxies. 
The numerical procedure employed is quite similar to that
used in Weil \& Hernquist (1993). 

Two interacting galaxies are initially separated by a rather large 
distance ($r_{\rm ini} = 5 =$ 75 kpc), so that tidal effects be negligible.
The initial velocities are taken as for parabolic encounter. 
The primary galaxy passes in a zero-inclination 
(in the plane of the gaseous disk of the donor galaxy -- $xy$ 
plane), prograde orbit (that is, the orbital angular momentum is parallel 
to the spin of the donor galaxy) with a pericenter distance 
$r_{\rm per}$ initially set as $1.6$. The polar axis of the ring-host 
galaxy disk component (parallel to the $x$-axis) lies in the orbital 
plane, so the orbit is polar regarding the ring-host galaxy.
The calculated orbit somewhat differs from parabolic (due to 
non-Keplerian potentials of galaxies), but its form is nearly the same 
for all models of a ring-host galaxy. 

\subsubsection{Results}

During the encounter, the primary galaxy strips the outskirts 
of the donor object and a ring, rotating in the direction of the 
orbital motion, eventually forms around the galaxy in the encounter 
plane. As the equatorial plane of the disk component of S0 is taken to 
be perpendicular to that of the orbital motion, the ring is polar. 
The total amount of accreted gas is about 10\% of all gas in the
donor galaxy (or about $2\times10^{9} M_{\odot}$) 
and there is not any significant difference in the stripped mass 
for all considered models of the ring-host galaxy. 
Some amount of the gas fell into the central part of galaxy. 
The mass of 
this gas is estimated as $2\times10^8 M_{\odot}$ within 1~ kpc from 
the center. The timescale of 
the ring formation is a few $10^8$ years. This time is somewhat shorter 
for rings forming around bulge-dominated galaxies and reaches up to 
$\sim 9 \times 10^8$ years for a disk-dominated model. The interacting
galaxies are separated by a sufficiently large distance (more than 
120 kpc) by the time of the ring steady-state settling, so that there are 
no any direct evidences of interaction between the galaxies and 
the ring-host system as an isolated object. 

The simulations have revealed an interesting feature which appears 
during the ring-forming process in different galaxian 
potentials. As a steady state sets in, ring sizes begin to diverge.
The results for four runs are presented 
in Figure~\ref{Frings}.  

\begin{figure*}
\picplace{0.8cm}
\caption{Time 10 (after the pericenter passage moment) frame for 4 runs. 
The values of the bulge to disk mass ratio of the ring-host galaxy are 
shown in the upper right-hand corner of all frames. All frames display 
the orbital plane ($xy$-plane) spatial projection and measure 30 kpc 
per edge. The equatorial plane of the ring-host galaxy disk component 
(parallel to $yz$-plane) is perpendicular to that of the orbital motion.} 
\label{Frings}
\end{figure*}   

Figure~\ref{Frings} gives a final stage of polar-ring formation for 
all considered models. This is the orbital plane projection of the 
gaseous rings. The marked difference in the ring structure is 
the difference in ring size, which rises when the 
bulge mass decreases and ranges from about 7 kpc in diameter 
for bulge-dominated systems (see Table 1) 
to approximately 13 kpc for disk-dominated 
objects.

\begin{figure}
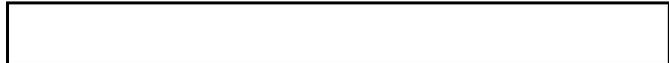

\picplace{0.8cm}
\caption{The angular momentum of a test particle on a circular orbit 
versus the distance from the center of a galaxy in the 
equatorial plane of the disk for 4 bulge-disk models. 
The short, medium and long dash lines conform to 
$M_{\rm b}/M_{\rm d}=2.0, 1.0, {\rm and\,} 0.5$, correspondingly; the 
solid line gives $v_{\rm rot}r-r$ relation for 
$M_{\rm b}/M_{\rm d}=0.2$. Two additional lines are shown as the 
bench-marks: ($*-*-*$) -- for diskless model, and 
($\circ-\circ-\circ$) -- diskless model with more concenrated bulge 
($a_{\rm b}=0.5$ kpc). The horizontal line marks the value of the mean 
angular momentum per ring particle for 4 runs.} 
\label{Fmom}
\end{figure}

The ring radius is finally determined by the angular momentum of 
donor galaxy gaseous particles forming the ring with respect to the 
galaxy at the moment of the pericenter passage. As the impact parameters 
are nearly the same for all runs, the value of the total angular momentum 
of ring-forming particles is almost the same for all models. But 
one can see from Figure~\ref{Fmom} that the positions of particles with 
the same angular momentum are different for galaxies with different 
structures. Particles 
are closer to the center of galaxy if there is a more 
concentrated mass distribution -- the bulge-dominated model. This implies, 
in particular, that under the same conditions rings forming around 
elliptical galaxies with a strong concentration of mass to the center 
(see, for instance, the curves marked as ($*-*-*$) and 
($\circ-\circ-\circ$) in 
Figure~\ref{Fmom}) should be less extended (on the 
average)~ --~ rather~ internal~ -- 
than those around galaxies with a more gently sloping 
density profile (this fact was first mentioned in Sotnikova 1996).
 
\begin{figure*}
\picplace{0.8cm}
\caption{Surface density profiles of rings. 
The short, medium and long dash lines conform to 
$M_{\rm b}/M_{\rm d}=2.0, 1.0, {\rm and\,} 0.5$, correspondingly; the 
solid line - $M_{\rm b}/M_{\rm d}=0.2$.} 
\label{F12}
\end{figure*}

In Figure~\ref{F12}, the surface density profiles along 
the major axes of the rings are shown. 
One can see one more interesting feature of simulated rings: rings 
farther from the center of the galaxy are more extended. 

These results give the post factum justification of our impact 
parameter choice. Indeed, a closer passage results in less extended 
rings. 
The angular momentum of a gaseous particle at the moment of 
the closest approach of galaxies with respect the ring-host 
system is significantly affected by the term 
proportional to $\sim v_{\rm per}r_{\rm per}$, 
where $v_{\rm per}$ is the ring-host galaxy velocity relative to the 
donor galaxy. For parabolic encounters, it leads 
to $\sim \sqrt{r_{\rm per}}$. As the ring size monotonically decreases, 
the angular momentum decreases; a closer encounter gives a 
ring lying  rather inside the galaxy. It is known that S0 galaxies 
possess some original gas, so the interaction between 
two gaseous systems would result in ring destruction. The further 
decreasing of pericenter distance leads to merging of the galaxies and 
total change of their structures. 
As to 
the more distant encounters, they do not allow capture of substantial 
amounts of gas, due to the sharp decrease of disturbing forces. We have 
repeated our runs with $r_{\rm per}=2.4=36$ kpc and have obtained that   
the mass of the resulting very diluted rings did not exceed 
$2\times10^8 M_{\odot}$. Hyperbolic
galaxy encounters are not efficient for 
ring formation, as the time of strong galaxy interaction is short. 
Let us note also that we discuss the properties
of classic PRGs with optical star-forming rings. So we discuss
the origin of relatively dense rings - with gas density large
enough for star formation. This gives an additional restriction on
the impact parameter, since the matter captured during the
distant encounter will spread along a more extended orbit
and will have lower mean density.  
Thus, galaxy encounters favourable to extended 
optical ring formation are extremely rare events. Such interactions
must be between galaxies of specific types and within
restricted geometry (polar encounters and a very narrow range of 
impact parameter). 

We found in our simulations that, with the $same$ impact parameter,
the rings forming around bulge-dominated galaxies are less
extended (by factor two) than the rings forming around disk-dominated
galaxies. Therefore, having the same accretion history in
the samples of elliptical and disk-type galaxies, we will
have {\it on average} more extended rings around disk-dominated
galaxies. This inference may explain 
in a quite natural manner the absence of extended polar rings 
around elliptical galaxies.
It can also explain a difference in morphology of ionized gas in
elliptical and S0 galaxies (Macchetto et al. 1996), assuming 
external origin of gas in these galaxies.

One can remark, however, that the obtained segregation of ring 
sizes (see~ Figure~\ref{Frings}) is not as clear as the observed 
dichotomy (see~ Table~ 1). The simple 
bulge-disk model giving internal rings for bulge-dominated objects 
fails to explain the existence of very extended (up to 30 kpc in diameter) 
polar rings around disk-dominated galaxies. Probably, there exists one 
more physical factor leading to the strong differences of the PRG 
structural properties.

\subsection{Exploration of dark halo inclusion effect}

As it was found from the observational data analysis, 
there exists for PRGs with extended rings 
a remarkable structural resemblence of host galaxies 
to late-type spiral galaxies. 
It is known that dynamical properties of 
late-type spiral galaxies are determined to a considerable extent 
by invisible massive halos (e.g., Freeman 1992). 
Hence, we can suppose that PRGs with extended rings possess a third 
global component (besides the bulge and the disk) - a dark massive halo. 
The existence of dark massive halos also follows from the analysis 
of ring kinematics (e.g., Schweizer et al. 1983, Reshetnikov \& Combes 1994a). 
To account 
for the rotational velocities in the rings of UGC 7576 and UGC 9796 inside 
radii of 17 kpc and 21.4 kpc, respectively, one requires amounts of dark mass 
reaching 1.6 and 3 times the luminous masses (Reshetnikov 
\& Combes 1994a).

What are the possible consequences of taking into account this 
structural component in the context of our investigation? As 
the gravitating mass of the primary galaxy increases, 
the orbital velocity of this galaxy relative to the donor object 
rises also. Then angular momentum arguments for a galaxy with 
such a structure lead to more distant orbits for captured 
particles. 

We changed our bulge-disk model 
by adding a smoothed third component - a halo. 
The structure and the shape of dark halos in early-type galaxies 
appears to be, at present, an open question 
(Sackett et al. 1994, Combes \& Arnaboldi 1996). For lack 
of unambiguous knowledge about the halo shape, we suppose the 
halo mass distribution to be spherical. As usual,  
halos are characterized by isothermal spheres over some radial 
interval. For simplicity, the folowing phenomenological cumulative 
mass profile is used to represent a halo in the present study:
\begin{equation}
M_{\rm h}(r) = \frac{M_{{\rm h}0}}{r_{\rm c}}\left(r-a_{\rm h} 
\arctan\left(\frac{r}{a_{\rm h}}\right)\right)
\left(1-e^{- \frac{r_{\rm c}}{r}}\right) \, ,\\
\label{halo}
\end{equation} 
where $M_{{\rm h}0}$ is the total halo mass, $r_{\rm c}$ a 
cutoff radius, and $a_{\rm h}$ a core radius. 

Our choice of halo parameters was determined by characteristic values 
obtained by Reshetnikov \& Combes (1994a) from the ring kinematics 
investigation. The core radius $a_{\rm h}$ was taken as 9 kpc. 
The value of the cutoff radius $r_{\rm cut}$ is somewhat 
arbitary and was taken to be equal to $3a_{\rm h}$. We choose 
this value to give a reasonable halo mass outside the radius comparable with 
$r_{\rm per}$. 
We have let the dark mass to be equal to twice
the luminous mass (bulge $+$ disk) inside a radius of 15 kpc.  
The total mass of the bulge and disk have been reduced, 
that is, $M_{\rm b} + M_{\rm d} = 0.5 = 5\times10^{10} M_{\odot}$, and 
we let $M_{\rm b}/M_{\rm d}=0.01$, $a_{\rm b} = 0.5$ kpc, 
$b_{\rm d} = 3$ kpc. 

The geometry of the 
encounter was the same as in previous numerical experiments. 
Expecting the capture of gaseous matter on distant orbits and 
formation of a more diluted object, we have increased the total 
number of particles (up to 20\,000) as well as their size $h$ 
up to 375 pc to keep the gas treatment as a continuous medium.    

The morphology of the new run is shown in 
Figure~\ref{Fhalo}. 
\begin{figure*}
\picplace{0.8cm}
\caption{Ring formation history for a galaxy with a massive halo. 
Dimensionless time (counted out from the moment of the minimum approach 
of galaxies) is given in the upper right-hand corner of all frames. 
The center of each frame coincides with the center of the ring-host 
galaxy. All frames show the orbital plane projection of the ring. The 
edge length of the frame corresponds to 80 kpc.} 
\label{Fhalo}
\end{figure*}   
By the time of $t=6$ after the pericenter 
passage, the captured gaseous matter is azimuthally 
smeared in a annular configuration around the galaxy. 
This structure gradually evolves into a closed ring. We 
have followed its evolution up to $t=20$. By this time, the 
ring was completly closed but not quite symmetric. 
According to Rix \& Katz (1991) for a ring forming 
as a result of the gaseous satellite disruption, the smearing process takes 
up to seven orbital periods - up to 3-4 Gyrs or two times 
the final time of our run. The total mass gathered in the 
ring-like structure does not significally differ from that obtained in 
haloless runs. The ring has a total mass of about 
$2.2\times10^9 M_{\odot}$. 
To explore the sensitivity of this value to the structural
properties of the donor galaxy, we have constructed an additional
model in which the spiral galaxy has its own halo. The parameters
of this halo were taken the same as for the host galaxy. The mass
of the captured gas turned out to be slightly smaller --
$1.4\times10^{9} M_{\odot}$.
(This value is somewhat small in comparison with 
that obtained from observations. We 
omit the discussion of this question, because this value is obtained 
for one set of orbital parameters and we did not investigate the role 
of initial conditions, impact parameter, and structure of the donor galaxy.
It may be impossible to understand the formation of very
massive rings in the frame of the pure accretion scenario and
the merging event is needed.)
 
The most remarkable feature of this 
annular structure is its size. The ring material lies well outside the 
luminous material of the host galaxy. 
We estimated the diameter of the ring 
of about 30 kpc. This value is typical of the objects of the first group 
of PRGs (see Table~ 1). 
The annular configuration obtained is rather narrow. 
Its further long-lived evolution will lead not only to azimuthal smearing 
of the gas but also to flattening of the radial density profile, 
forming a disk-like structure. Two factors promote the formation
of rings with extended density distributions - the viscosity
(see Rix \& Katz 1991) and the nonsphericity of the potential
(see Fig.5 in this paper and Fig.1 in Katz \& Rix 1992). 

\section{Discussion and conclusions}

We have analyzed the data on global structural properties of all 
known PRGs, and came to a conclusion that there is a correlation 
between the main characteristics of polar rings and host galaxies. 
There exist two different classes of PRGs: bulge-dominated 
galaxies with internal (relative to the optical size of a galaxy) 
rings and disk-dominated systems with extended rings. 
The sample studied is somewhat small (8-9 objects), but when we 
combine our preliminary inference with the results of 
numerical experiments, a plausible picture of PRG properties 
emerges.   

We have presented the investigation of the ring-forming process by 
studying the stripping of gas-rich galaxy outskirts during the 
encounter with galaxies of different structures and the subsequent 
settling of the captured material in the potential of a companion 
galaxy using an SPH method. 
The simulations have demonstrated that the ring sizes depend on 
galaxy potentials. 
{\it Under the same interaction conditions, rings twisting around 
galaxies with a strong concentration of mass to the center 
(bulge-dominated systems) are less extended (on the average) than 
that around disk-dominated galaxies.} 
Moreover, we have found that it is impossible to obtain 
very extended rings (with a diameter of about 30 kpc) 
taking into account the distribution of the luminous mass only. 
To obtain such extended rings, the galaxy mass distribution  
should be less concentrated than observed luminosity distribution. 
Therefore the presence of dark halos is required to explain the 
existence of extended rings.

One can suppose that the observational
dichotomy is the result of selection. Indeed, interactions between
galaxies accompanied by matter accretion occur with participation
of various types of galaxies. But we have more chances to create
long-lived extended rings when the accreting galaxy is a gas free, 
disk-like galaxy
with extended massive dark halo. So such extended polar rings 
are expected to be observed
preferably around specific types
of galaxies.

The existence of the
observational dichotomy suggests that PRGs with bulge-dominated
central galaxies may possess less
pronounced dark haloes in comparison with the first group of objects
(another possibility - bulge-dominated galaxies have more
centrally-concentrated halos - does not change our results as 
follow from Sect.3.1.3). These conclusions do not contradict the present-day 
observational data about dark halos in early-type galaxies (e.g., 
Bertin \& Stiavelli 1993, de Zeew 1995). 
Nevertheless, we do not exclude other explanations of the revealed
observational dichotomy (for instance, environmental influence etc.).

As was mentioned in Section 2, there is a close similarity 
between extended rings and disks of late-type spiral galaxies. 
Extended rings have the same total 
luminosities, integral colors, ratios of HI mass to the blue luminosity 
corrected for internal absorption, gas-to-dust 
ratios as normal spiral galaxies; 
there is evidence for the gaseous medium in polar 
rings to be near (or above) the gravitational stability limit, as 
the gaseous disks in spiral galaxies (Reshetnikov et al. 1994). 
Two polar rings (in UGC~ 7576 and UGC~ 9796) demonstrate 
H$\alpha$-derived global star formation rates to be typical 
for normal spiral galaxies; general properties of HII regions 
in these galaxies are normal for late-type spirals 
(Reshetnikov \& Combes 1994a). HII regions in the ring of
prototype polar-ring galaxy NGC~ 2685 demonstrate nearly
solar oxygen abundances (Eskridge \& Pogge 1994).
Extended rings show radial color
gradients with bluer colors at larger radii (Reshetnikov et al. 1994,
Arnaboldi et al. 1995). The HI/$\rm H_{2}$ mass ratio in the rings
of NGC~ 660 and NGC~ 4650A is usual for spiral galaxies 
(Combes et al. 1992, Watson et al. 1994). There is evidence for  
the spiral structure in the rings of UGC~ 7576 
(Reshetnikov \& Combes 1994b) and NGC~ 4650A
(Arnaboldi et al. 1996). There is an indication of the presence 
of a large-scale magnetic field in the ring of NGC~ 660 
(Reshetnikov \& Yakovleva 1991). Moreover, according to 
Combes \& Arnaboldi (1996) and Arnaboldi et al. (1996), dark matter 
in PRGs could co-exist with the HI component, leaving its mark on 
the kinematics of the polar ring itself similar to the situation 
with a late-type galaxy gaseous disk. One can note finally that PRGs 
lie close to the Tully-Fisher relation 
for spiral galaxies (Knapp et al. 1985). Therefore, PRGs with extended 
rings could be undistinguishable from normal spiral galaxies at 
less advantageous orientation (with a more face-on ring) - see the 
discussion of this question in PRC. Probably, they will look like 
early-type spiral galaxies (maybe barred) with extended low surface 
brightness disks. 

The striking similarity between extended rings and spiral disks  
suggests that PRGs $are$ giant low surface-brightness spiral 
galaxies with decoupled bulges. Although this cannot be completely 
ruled out in some special cases, we suppose that the ring capture 
due to external accretion is a more frequent process 
(see Section 1 about observations of forming rings). 
Therefore, in contrast to the usual opinion that galactic 
interactions lead to fast morphological changes towards S0/E, 
the observational data on PRGs and our modelling simulations 
provide the example of a rare opposite shift - from an early-type 
gas-free galaxy through the capture of a gaseous disk to a spiral galaxy. 
One can suppose that PRG formation due to external accretion 
can imitate in some aspects the formation of disk galaxies in 
the hierarchically clustering model of the Universe (e.g., Steinmetz 1996). 

In conclusion, let us reply shortly to the main questions of the article 
(see Section 1):

the accretion of matter from a gas-rich companion is a 
quite effective mechanism for polar-ring formation;

the spatial size of the forming ring is determined (under the same 
interaction conditions) by the mass distribution of the galaxy;

the process of ring formation takes approximately 
$\rm (7-9)~\times~10^{8}$ years for haloless systems and reaches up to  
a few Gyrs for host-galaxies possessing massive halos; 

the total mass captured into the ring 
during an encounter with a gas-rich spiral galaxy exceeds 
$10^{9}\,M_{\odot}$.

\acknowledgements
{We would like to thank Fran\c coise Combes and the referee
(Linda Sparke) for helpful comments
and criticism. VR acknowledges support from French 
Minist$\grave{\rm e}$re de la Recherche et de la Technologie during his stay
in Paris. This work was supported by grant $N$ 94-02-06026-a from Russian 
Foundation for Basic Research.}

\end{document}